\documentclass[aps,showkeys,preprint]{revtex4}%
\usepackage{amsfonts}
\usepackage{amsmath}
\usepackage{amssymb}
\usepackage{graphicx}%

\begin{document}
\title{Analysis of the Fusion Hindrance in Mass-symmetric Heavy Ion Reactions}
\author{Caiwan Shen$^{1}$, Yasuhisa Abe$^{2}$, Qingfeng Li$^{1}$, David Boilley$^{3,4}%
$}
\affiliation{$^{1}$School of Science, Huzhou Teachers College, Huzhou 313000, China}
\affiliation{$^{2}$Research Center for Nuclear Physics, Osaka University, Ibaraki (Osaka),
567-0047 Japan }
\affiliation{$^{3}$GANIL, BP 55027, Caen cedex 5, F-14076 Caen, France}
\affiliation{$^{4}$ University of Caen, B.P. 5186, F-14032 Caen Cedex, France }
\begin{abstract}
The fusion hindrance, which is also denominated by the term extra-push, is
studied on mass-symmetric systems by the use of the liquid drop model with the
two-center parameterization. Following the idea that the fusion hindrance
exists only if the liquid drop barrier (saddle point) is located at the inner
side of the contact point after overcoming the outer Coulomb barrier, the
reactions in which two barriers are overlapped with each other are determined.
It is shown that there are many systems where the fusion hindrance does not
exist for the atomic number of projectile or target nucleus $Z\leq43$, while
for $Z>43$, all of the mass-symmetric reactions are fusion-hindered.

\end{abstract}
\maketitle

\section{Introduction}

Nuclear reactions with heavy-ion beams have been revealing many interesting
behaviours of the atomic nucleus. Among them, fusion of two heavy ions into
one spherical nucleus is an interesting process which is not yet well
understood, although the fusion probability is crucially important for
synthesis of the so-called super-heavy elements. It is well known that fusion
reactions of light systems and heavy systems have different feature: in light
systems, the fusion cross section can be explained by the overcoming of the
Coulomb barrier between projectile and target (for instance, see Ref.
\cite{exp-no-ep1,exp-no-ep2}), while in heavy systems, the experimental fusion
cross sections start to be appreciable at higher energies than the Coulomb
barrier, that is, are much smaller than those calculated with the same model
as that of the light systems (examples at Ref. \cite{exp-ep1,exp-ep2}). It
means that the fusion barrier in heavy systems appears to be higher than the
Coulomb barrier. The phenomenon in heavy systems is called fusion hindrance,
and the corresponding energy difference between the fusion and the Coulomb
barriers is called extra-push energy. Theoretical attempts to explain this
phenomenon are made with moderate success. In Ref.
\cite{swiatecki1,royer,swiatecki2}, it was explained by an internal barrier
which must be overcome after passing over the usual Coulomb barrier. The
internal barrier could be thought as the conditional saddle point in the
liquid-drop potential as well as could be attributed to an effective barrier
due to the dissipation of the incident kinetic energy\cite{gross, frobrich}.
But there was no simple explanation of the mechanism of the hindrance, and 
therefore no theoretical predictions with quantitative reliability.

Since there are two barriers for the fusion, we have proposed a model where
the fusion reaction is divided into two steps: (i) the projectile and target
overcome the Coulomb barrier and reach the contact configuration, (ii) the
touched projectile and target evolve from di-nucleus to the spherical compound
nucleus by passing over the ridge line of the LDM potential \cite{abe-eurojp}.
The model explains the extra-push energy and furthermore gives an energy
dependent fusion cross sections \cite{shen2002, PTPSUPPL, shen2008}. According
to the compound nucleus theory, production cross sections are given by the
fusion and the survival probabilities.

In the present paper, presuming that the internal barrier plays a crucial role
in the fusion hindrance, we analyze a relation between the saddle point (more
generally a conditional saddle or a ridge line) and the occurrence of the
hindrance. The dissipative dynamics of the passing-over of the saddle point
has been already studied analytically with the simplification of the inverted
barrier \cite{abe2000}. According to the results of Ref. \cite{abe2000}, the
hindrance is given by the saddle point height measured from the energy of the
di-nucleus configuration formed by the projectile and the target nuclei of the
incident channel. In other words, there is no hindrance for cases with no
saddle point height. Therefore, the border between the normal and the hindered
fusion is given by the condition that the saddle point height be equal to
zero, that is, the di-nucleus configuration be on the top of the saddle point.

In mass-symmetric reactions, the ridge line is simplified into a saddle point,
which makes analysis to be simpler. Below, we will find out the region of
fusion hindrance for mass-symmetric reactions by using the finite range LDM
with two-center parametrization of nuclear shapes \cite{sue, u-ldm-sato}.

Experimentally, it is very difficult to distinguish between fission events
coming from the fused compound nuclei and so-called quasi-fission events
coming from a di-nuclear system. Therefore, experimental fusion cross sections
might not be reliable enough for quantitative comparisons with theoretical
calculations. In the present paper, we, thus, focus on the appearance and
disappearance of the hindrance that is clearly observed in the symmetric systems.

The present paper is organized as follows: Sec. II recapitulates the
parametrization of the di-nuclear system. The determination of the neck
parameters which has been recently obtained by the present authors
\cite{stras, F-J, boilley} is reminded, and a prediction of the fusion
hindrance area is shown in Sec. III for mass-symmetric systems. Sec. IV gives
a summary.

\section{Parametrization of di-nuclear system}

\begin{figure}
[ptbh]
\begin{center}
\includegraphics[
width=2.5371in
]%
{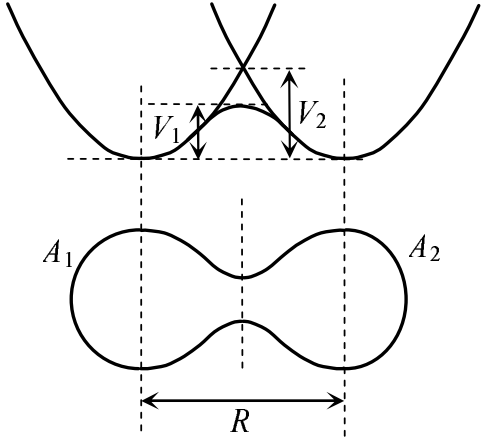}%
\caption{Schematic plot of the parametrization of a di-nuclear system. The
upper figure shows the harmonic oscilator potential of projectile and target,
while the lower one shows the cross section of an equi-potential surface. See
text for details.}%
\label{fig1}%
\end{center}
\end{figure}

There are several ways to parametrize the shape of the amalgamated system. The
more accurate description, the more parameters. In this paper we use the
two-center parametrization, using three important parameters which are:
distance between two centers $z$, the mass asymmetry parameter $\alpha$, and
the neck parameter $\varepsilon$, as shown in Fig. \ref{fig1}. The first one
is defined as a dimensionless parameter as follows,%
\[
z=R/R_{0},
\]
where $R$ denotes the distance between two centers of the harmonic potentials,
and $R_{0}$ the radius of the spherical compound nucleus. The mass-asymmetry
parameter is defined as usual,
\[
\alpha=\frac{A_{1}-A_{2}}{A_{1}+A_{2}},
\]
where $A_{1}$ and $A_{2}$ are mass numbers of the constituent nuclei. The neck
parameter $\varepsilon$ is defined by the ratio of the smoothed height at the
connection point of the two harmonic potentials ($V_{1}$) and that of spike
potential ($V_{2}$), i.e.,
\[
\varepsilon=V_{1}/V_{2}.
\]
In this description, nuclear shape is defined by equi-potential surfaces with
a constant volume. For example, $\varepsilon$ = 1.0 means no correction, i.e.,
complete di-nucleus shape, while $\varepsilon$ = 0.0 means no spike, i.e.,
flatly connected potential, which describes highly deformed mono-nucleus.
Thus, the neck describes shape evolution of the compound system from
di-nucleus to mono-nucleus. The initial parameters for $z$ and $\alpha$ are
\[
z_{0}=\frac{A_{p}^{1/3}+A_{t}^{1/3}}{(A_{p}+A_{t})^{1/3}},
\]
and
\[
\alpha_{0}=\frac{A_{t}-A_{p}}{A_{t}+A_{p}},
\]
respectively. In mass-symmetric case, $z_{0}=\sqrt[3]{4}=1.5874$ and
$\alpha=\alpha_{0}=0.$ The initial value of $\varepsilon$ will be explained in
the next section.

\section{Calculations and analysis}

\subsection{Determination of the neck parameter}

\begin{figure}
[ptb]
\begin{center}
\includegraphics[
width=4.8634in
]%
{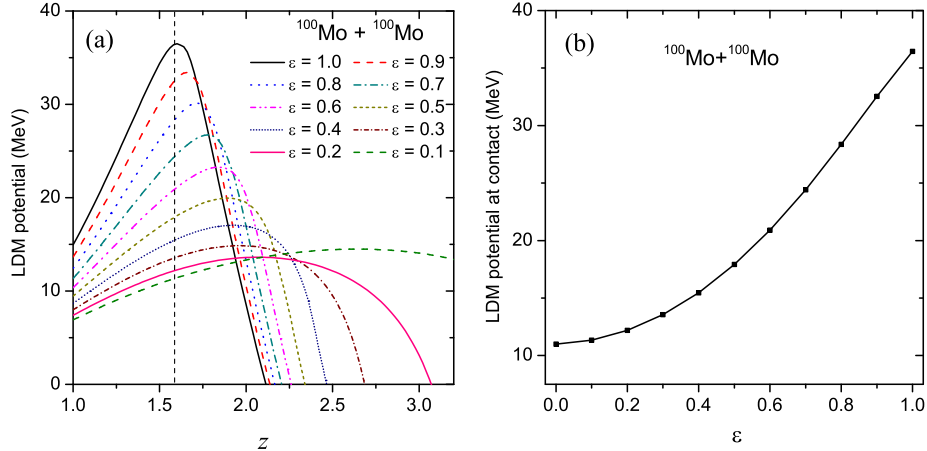}%
\caption{LDM potential for $^{100}$Mo + $^{100}$Mo. Left panel: the potential
as a function of $z$ with various $\varepsilon$, which gives the $\varepsilon
$-dependence of the saddle point. The vertical dashed line represents the
contact point. Right panel: the potential at contact configuration with
respect to $\varepsilon$, which is used to calculate the average of
$\varepsilon$.}%
\label{fig2}%
\end{center}
\end{figure}

With the parametrization of the amalgamated system, the finite range LDM
potential can be calculated. In order to study the neck-dependence of the
saddle point, the LDM potentials of, as an example, $^{100}$Mo + $^{100}$Mo as
a function of $z$ for different neck parameters are given in Fig.
\ref{fig2}(a). It is clearly shown that the saddle point is very sensitive to
the $\varepsilon$: when $\varepsilon$ is smaller (thicker neck), the saddle
point is shifted to lower and wider place. Therefore, how the neck changes at
contact configuration is a very important problem. To reveal the driving
effect of the LDM potential on the neck, the relation between LDM potential at
contact configuration and $\varepsilon$ is plotted in Fig. \ref{fig2}(b). The
large positive slope of LDM potential with respect to $\varepsilon$
($dV/d\varepsilon$) drives the neck at contact to be thicker with
$\varepsilon$ up to 0. This is natural, considering the strong surface tension
of the nuclear matter and a sensitive change of the surface area due to the
variation of the $\varepsilon$. Since we also know that the inertia mass for
the $\varepsilon$ degree of freedom is small, its momentum is expected to be
quickly equilibrated, compared with the other two degrees of freedom: so
$\varepsilon$ very quickly reaches the end at $\varepsilon$ = 0.0, starting
with $\varepsilon$ = 1.0. Actually, due to actions of the random force
associated to the friction, the $\varepsilon$ reaches the equilibrium quickly,
far quicker than the time scale of the radial fusion motion \cite{stras, F-J,
boilley}. Thus, when the projectile and target touch with each other, the neck
firstly reaches its equilibrium, and then the other two degrees of freedom
start to evolve toward the compound stage. The neck parameter at equilibrium
can be determined through the average of $\varepsilon$ via%
\[
\left\langle \varepsilon\right\rangle =\int\varepsilon w(\varepsilon
)d\varepsilon/\int w(\varepsilon)d\varepsilon,
\]
where $w(\varepsilon)=e^{-V(\varepsilon)/T}$, and $T$ is the temperature of
the system. In most cases, $\left\langle \varepsilon\right\rangle $ is close
to 0.1. Therefore we take $\varepsilon=0.1$ in next calculations. This value
is also used in the fusion cross section calculations in two-step model
\cite{shen2008}, which shows a good agreement with experimental data.

From Fig. \ref{fig2}(a), it is obvious that, for $^{100}$Mo + $^{100}$Mo, the
contact point $z_{0}$ $(=1.5874)$ is located inside the saddle point
$z_{\text{saddle}}$ $(=2.40)$ at $\varepsilon=0.1$, which means that the
di-nucleus automatically reaches the compound nucleus after overcoming the
Coulomb barrier, i.e., no fusion hindrance exists for this case. Following the
same way, we can make the same calculation for each mass-symmetric reactions
to find out the region where the fusion hindrance disappears, or extra-push
energy is zero.

\subsection{Fusion hindrance region}

To study the appearance of the hindrance phenomena, one should compare the
relative positions of the Coulomb barrier and the conditional saddle point
with the neck set at $\varepsilon=0.1$. The location of the Coulomb barrier
depends on the model, but also on the neck parameter \cite{Iwa}, but it is
always beyond the contact point of the two rigid spheres at contact. Here,
with a conservative point of view, we will choose this contact point as a reference.

For a certain mass-symmetric reaction, the conditional saddle point position
can be compared with the contact point to determine if the extra-push appears
or not. If the contact point is located on the inner side of saddle point, the
di-nucleus system will evolve automatically from the touching point to the
compound stage by the driving force $dV_{\text{LDM}}/dz$. Otherwise, an
additional LDM barrier has to be overcome, which needs an extra-push energy.%
\begin{figure}
[ptb]
\begin{center}
\includegraphics[
width=2.8618in
]%
{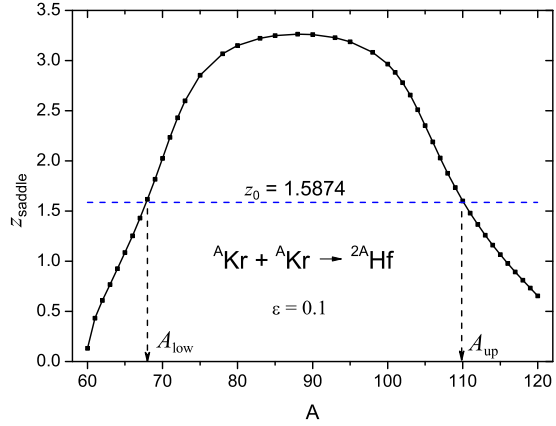}%
\caption{Determination of the region where the fusion hindrance disappers for
mass-symmetric reactions $^{A}Z+$ $^{A}Z\rightarrow$ $^{2A}(2Z)$. If the
distance $z$ at saddle point is larger than the $z_{0}$ ($A_{\text{low}}\leq
A\leq A_{\text{up}}$), fusion hindrance of the reaction does not exist.
Otherwise, the extra-push energy is needed to overcome the LDM saddle. The
dashed horizontal line represents the contact point.}%
\label{fig3}%
\end{center}
\end{figure}

To find out the fusion hindrance region of the reaction $^{A}Z+$
$^{A}Z\rightarrow$ $^{2A}(2Z)$, we fix $Z$ and determine the saddle point for
each $A$. Fig. \ref{fig3} shows an example for $Z=36$. For very small and very
large $A$ the contact point is located outside the saddle point, while for
$A_{\text{low}}\leq A\leq A_{\text{up}}$, the contact point is inside the
saddle point, which means that there is no fusion hindrance in this region.
Therefore, $A_{\text{low}}$ and $A_{\text{up}}$, where the contact point is
overlapping with the saddle point, are critical mass numbers\ corresponding to
the $Z$.%
\begin{figure}
[ptb]
\begin{center}
\includegraphics[
width=3.0228in
]%
{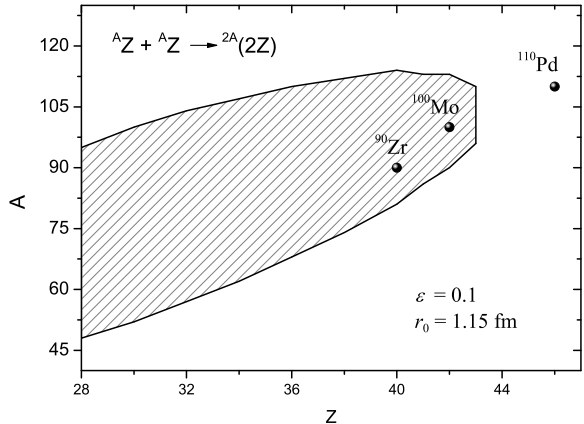}%
\caption{Critical line for mass-symmetric reactions where the contact point is
overlapped with the saddle point. In the shadowed region the contact point is
located inside the saddle which means that there is no fusion hindrance in the
corresponding reaction. While in white area it is on the contrary and the
extra-push energy is needed. Three dots correspond to $^{90}$Zr + $^{90}$Zr,
$^{100}$Mo + $^{100}$Mo and $^{110}$Pd + $^{110}$Pd reactions. In the
calculations, $r_{0}=1.15$ fm and $\varepsilon=0.1$ are adopted.}%
\label{fig4}%
\end{center}
\end{figure}

Changing the proton number $Z$, series of $A_{\text{low}}$ and $A_{\text{up}}$
are determined and plotted in Fig. \ref{fig4}. The shadowed area (including
the border) represents the reactions of $A_{\text{low}}\leq A\leq
A_{\text{up}}$ and consequently where fusion hindrance does not exist, while
the white space represents the contrary. It is interesting that for $Z<42$,
both $A_{\text{low}}$ and $A_{\text{up}}$ increase with increasing $Z$, but
the width of ($A_{\text{up}}-A_{\text{low}}$) becomes narrower and narrower.
When $Z$ is larger than 43, all of the LDM saddle points of the reaction
$^{A}Z+$ $^{A}Z$ are located inside the contact points. Therefore, the
extra-push energy should be considered for all mass-symmetric reactions with
$Z>43$. It is well known that the reactions $^{90}$Zr + $^{90}$Zr and $^{100}%
$Mo + $^{100}$Mo do not have fusion hindrance while the $^{110}$Pd + $^{110}%
$Pd reactions does \cite{keller,schmidt,morawek}. To compare with the above
theoretical analysis, the three reactions are also pointed in Fig. \ref{fig4},
in which two dots for $^{90}$Zr and $^{100}$Mo systems are inside the shadowed
area, while the dot for $^{110}$Pd system is in the white space. The results
show that the theoretical analysis is in a good agreement with experimental
data for mass-symmetric reactions. Using $Z^{2}$ as a criteria, $Z^{2}>1849$,
which is also in agreement with the empirical rule used to determine the
appearance of the hindrance.

\section{Summary}

In summary, the fusion hindrance of mass-symmetric reactions is studied with
the two-center model, in which three parameters (dimensionless distance
between two centers $z$, neck parameters $\varepsilon$, mass asymmetry
parameter $\alpha$) are employed to describe the di-nuclear system. Because of
the very fast evolution of $\varepsilon$ compared to the other two degrees of
freedom, $\varepsilon$ is set to its equilibrium value 0.1. In order to find
out the reactions where the fusion hindrance does not exist, the position of
the saddle point and the contact point for $^{A}Z+$ $^{A}Z$ are compared for
different $Z$ and $A$. It is found that the mass-symmetric fusion reactions
without hindrance are located only in a limited area, and $Z$ should be $\leq$
$43$. While for systems with $Z$ larger than 43, all of the mass-symmetric
reactions are hindered, i.e., the extra-push energy should be required to form
the compound nucleus.

Experimental studies around the predicted border are strongly called for.
Quantitative comparisons of fusion cross sections or of fusion probabilities
should be made between experiments and theoretical results, for which the
sticking probability in the two-step model, i.e., effects of over-coming of
the Coulomb barrier have to be taken into account, though there are ambiguities.

Following the same method, the fusion hindrance in mass-asymmetric reactions
can also be studied. However, the determination of the saddle in two
dimensional LDM potential is more complicate than the symmetric case. Studies
along this direction are currently underway and will be addressed in a
forthcoming paper.

\begin{acknowledgments}
The present work has been supported by Natural Science Foundation of China
under grant No. 10675046, the key project of the Ministry of Education of
China under grant No. 209053 and by JSPS grant No. 18540268. The authors also
acknowledge supports and hospitality by RCNP Osaka University, GANIL, and
Huzhou Teachers College, which enable us to continue the collaboration. Two of
the authors (C.S., Q.L.) are also very grateful to Professor En-Guang Zhao for
helpful discussions and encouragements.
\end{acknowledgments}

\end{document}